\documentclass[12pt]{article}
\usepackage{graphicx}

\newcommand\pubnumber{SNSN-323-63}
\newcommand\pubdate{\today}

\def\institute{Institute for Theoretical Physics\\
Heidelberg University, D-69120 Heidelberg, GERMANY}

\def\Title#1{\begin{center} {\Large #1 } \end{center}}
\def\Author#1{\begin{center}{ \sc #1} \end{center}}
\def\Address#1{\begin{center}{ \it #1} \end{center}}

\newcommand\pubblock{\rightline{\begin{tabular}{l} \pubnumber\\
         \pubdate  \end{tabular}}}
\newenvironment{Abstract}{\begin{quotation}  }{\end{quotation}}
\newenvironment{Presented}{\begin{quotation} \begin{center} 
             PRESENTED AT\end{center}\bigskip 
      \begin{center}\begin{large}}{\end{large}\end{center} \end{quotation}}
\def\Acknowledgements{\bigskip  \bigskip \begin{center} \begin{large}
             \bf ACKNOWLEDGEMENTS \end{large}\end{center}}

\def\beq{\begin{equation}}
\def\eeq#1{\label{#1}\end{equation}}
\def\eeqn{\end{equation}}

\def\beqa{\begin{eqnarray}}
\def\eeqa#1{\label{#1}\end{eqnarray}}
\def\eeqan{\end{eqnarray}}

\let\bar=\overbar

\def\Dslash{\not{\hbox{\kern-4pt $D$}}}
\def\dslash{\not{\hbox{\kern-2pt $\del$}}}

\def\msb{{\bar{\ssstyle M \kern -1pt S}}}

\begin{document}
\begin{titlepage}
\pubblock

\vfill
\Title{Global Searches for New Physics with Top Quarks}
\vfill
\Author{Susanne Westhoff}
\Address{\institute}
\vfill
\begin{Abstract}
\noindent This is a brief summary of the latest searches for virtual effects of new physics in the top sector. In the framework of the Standard Model Effective Field Theory (SMEFT), I show how to resolve the structure of effective couplings by combining observables at the LHC and at flavor experiments in a global fit. With this approach we start exploring the features of a UV theory at energies beyond current colliders.
\end{Abstract}
\vfill
\begin{Presented}
$13^\mathrm{th}$ International Workshop on Top Quark Physics\\
Durham, UK (videoconference), 14--18 September, 2020
\end{Presented}
\vfill
\end{titlepage}
\def\thefootnote{\fnsymbol{footnote}}
\setcounter{footnote}{0}
%

\section{Introduction}
\noindent In our search for new physics, the top quark is interesting in two respects. At the LHC, direct top production allows us to probe virtual effects of heavy new particles at high energies and in various precision observables. At flavor experiments, the top quark induces meson decays and mixing, which are rare in the Standard Model (SM) and therefore very sensitive to new physics. In combination, both areas allow us to resolve fundamental interactions at energies above the reach of the LHC.

Assuming that new interactions, if present, preserve the SM gauge symmetries, we can consider the Standard Model as an effective field theory, SMEFT, that describes the low-energy appearance of a fundamental theory at high energies. This framework allows us to classify effects of new physics at scales $\Lambda > 1$\,TeV in sub--TeV observables in terms of effective couplings $C_i$ called Wilson coefficients. Based on the effective Lagrangian
\beq
\mathcal{L}_{\rm SMEFT} = \mathcal{L}_{\rm SM} + \frac{C_5}{\Lambda}O_5 + \sum_i \frac{C_i}{\Lambda^2} O_i + \dots\,,
\eeqn
we calculate LHC observables as polynomials of SMEFT coefficients
\beq
\sigma = \sigma_{\rm SM} + \sum_i\frac{C_i}{\Lambda^2}\sigma_i + \sum_{i,j}\frac{C_iC_j}{\Lambda^4}\sigma_{ij}\,,
\eeqn
which can be compared with data. In a concrete model, UV physics imprints itself onto certain Wilson coefficients, which leave a pattern of effects in observables at lower energies. Most models generate effects of several Wilson coefficients in several observables. To resolve the SMEFT parameter space and pin down the features of the UV theory, a global analysis is required.

In the top sector, several global SMEFT analyses of LHC data have demonstrated sensitivity to new physics above, but not far above the TeV scale~\cite{Buckley:2015lku,Hartland:2019bjb,Brivio:2019ius,Durieux:2019rbz}. The main asset of the top sector, however, is not the reach in energy, but the large number of precise observables, which allow us to resolve much of the set of relevant Wilson coefficients. In Sec.~\ref{sec:top}, we will combine precise observables in top-antitop production resolve the chiral structure of effective four-quark couplings.

Combining LHC observables with flavor observables greatly improves the sensitivity to new physics and probes blind directions in global fits of LHC data~\cite{Bissmann:2019gfc,Aoude:2020dwv,Bissmann:2020mfi,Bruggisser:2021duo}. Below the scale of electroweak symmetry breaking, interactions of light fermions, the photon and the gluon are well described by the Weak Effective Theory (WET)
\beq
\mathcal{L}_{\rm WET} =  \mathcal{L}_{\rm QED} + \mathcal{L}_{\rm QCD} + \sum_a \mathcal{C}_a \mathcal{O}_a + \dots\,.
\eeqn
The set of Wilson coefficients $\{\mathcal{C}_a\}$ in WET is different from and smaller than the set $\{C_i\}$ in SMEFT, mainly because the top quark and the weak bosons are no longer dynamical constituents of the theory. Both sets are connected through the renormalization group. The respective effects in high-energy and low-energy observables, however, depend on the assumed flavor structure of the SMEFT coefficients. In Sec.~\ref{sec:top-bottom}, we will show how to resolve the flavor structure of a UV theory in a combined analysis of top and bottom observables.

\section{New directions in the top sector}\label{sec:top}
\noindent To illustrate how to probe the features of new physics interactions in top observables, we focus on two SMEFT operators
\beq
O_{tq}^8 = (\bar{t}_R\gamma_\mu T^A t_R)(\bar{q}_L \gamma^\mu T^A q_L),\quad O_{Qq}^{1,8} = (\bar{Q}_L\gamma_\mu T^A Q_L)(\bar{q}_L \gamma^\mu T^A q_L)\,,
\eeqn
where $q = (u_L,d_L)$ and $Q = (t_L,b_L)$ stand for weak doublets of left-handed quarks from the first two and from the third generation, respectively. These two operators only differ by the chirality of the heavy quarks, i.e., $t_R$ versus $t_L$ for the tops. Distinguishing between the effects of $O_{tq}^8$ and $O_{Qq}^{1,8}$ in observables means probing the chiral structure of a UV theory that induces them. This example is                                                                                                                                                                                                                                                                                                                                                                                              part of a comprehensive analysis performed in Ref.~\cite{Brivio:2019ius}, to which we refer you, dear interested reader, for details.

\begin{figure}[t!]
\centering
\includegraphics[width=0.48\textwidth]{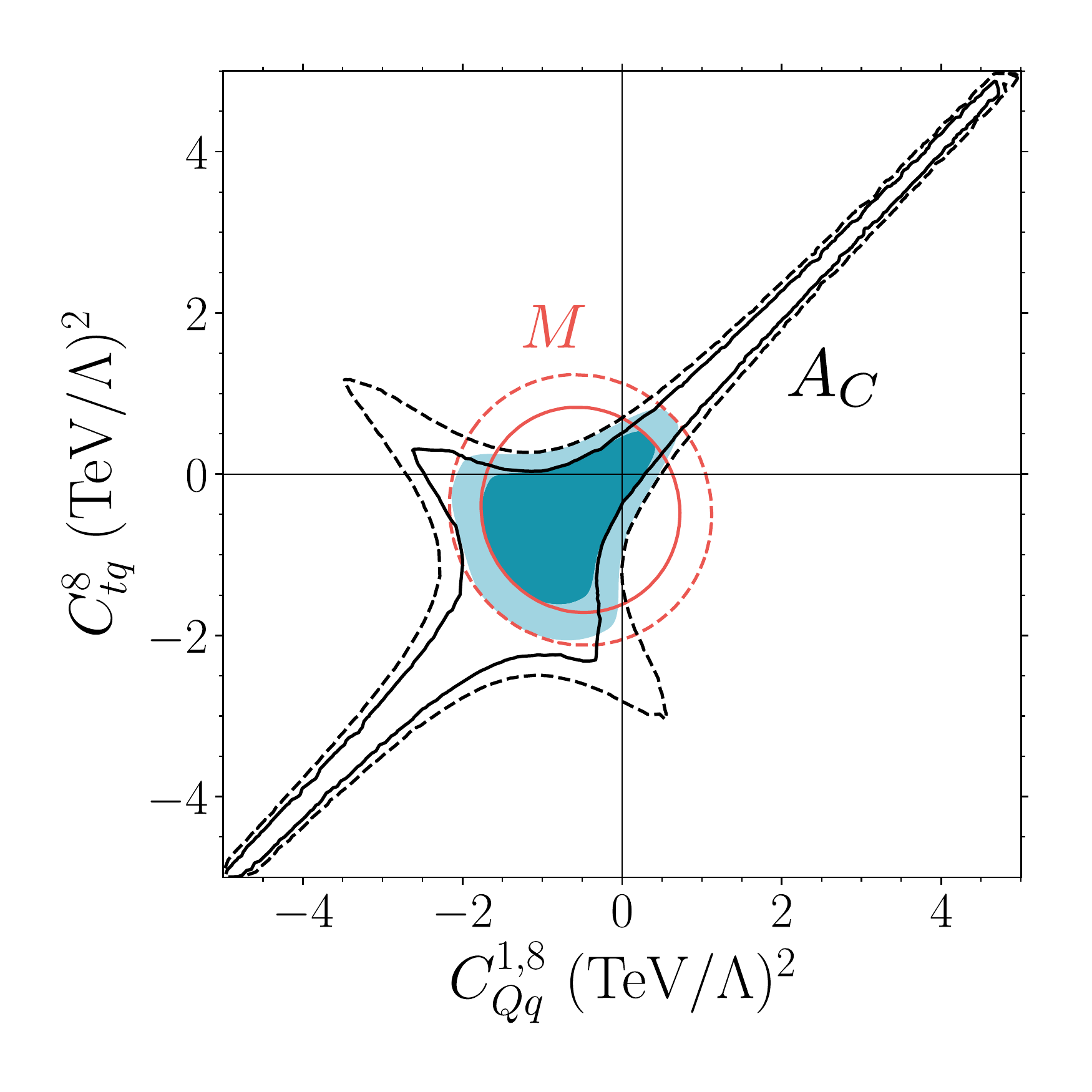}\hspace*{0.02\textwidth} \includegraphics[width=0.48\textwidth]{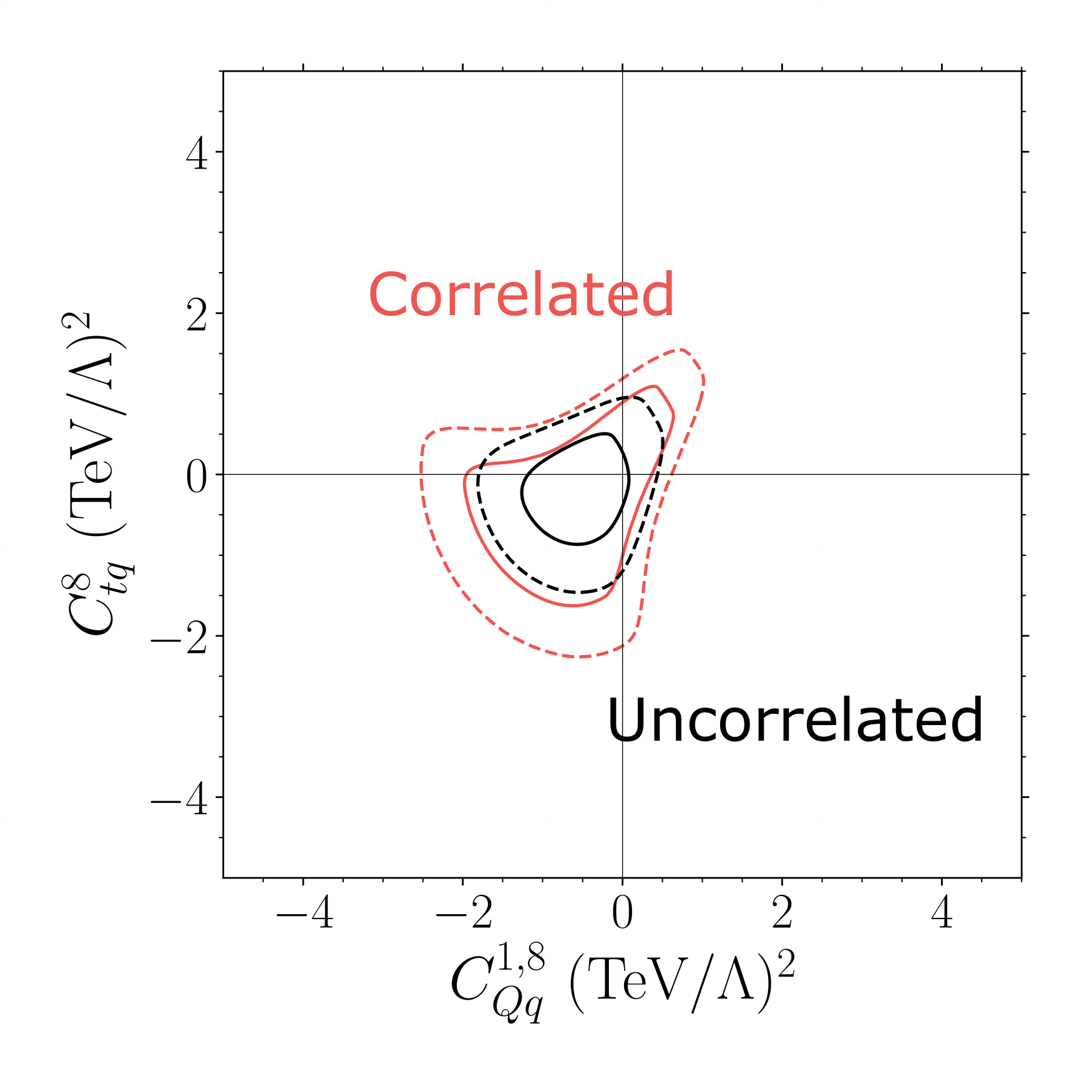} 
\caption{Left: bounds on the chiral four-quark couplings $C_{Qq}^{1,8}$ and $C_{tq}^8$ obtained from a combined fit to charge-symmetric $t\bar t$ observables (red), charge asymmetries (black), and both sets of observables combined (blue)~\cite{Brivio:2019ius}. Right: impact of correlations in a combined fit of $t\bar t$ observables~\cite{correlations}; shown are the bounds obtained by assuming that systematic uncertainties in bins of differential distributions are uncorrelated (black) and 100\% correlated (red). Solid and dashed lines mark the Gaussian equivalent of $\Delta \chi^2 = 1,4$.}
\label{fig:top-antitop}
\end{figure}

Top-antitop production is a good test ground for chiral top-quark couplings. Charge-symmetric observables like the total cross section probe mostly vector-like $L+R$ couplings, while charge asymmetries probe axial-vector couplings $L-R$. In terms of Wilson coefficients, they read
\beqa
\sigma_{t\bar t} & = & \sigma_{\rm SM} + \sigma_{VV}\big(C_{Qq}^{1,8} + C_{tq}^8\big) + \mathcal{O}(C^2)\\\nonumber
A_C & = & \left(\sigma_{\rm SM}^A + \sigma_{AA}\big(C_{Qq}^{1,8} - C_{tq}^8\big)\right)/\sigma_{t\bar t} + \mathcal{O}(C^2)\,.
\eeqan
Quadratic contributions of $\mathcal{O}(C^2)$ have been neglected here, but are numerically relevant when comparing the predictions with data. In Fig.~\ref{fig:top-antitop}, left, we show the results of fits to LHC top data with charge-symmetric observables (red contours), charge asymmetries (black contours), and the combination of both sets (blue areas). The hyperbola-shaped bounds obtained from the asymmetries leave a blind direction along $C_{Qq}^{1,8} + C_{tq}^8$, which is broken when adding charge-symmetric observables. This simple example demonstrates how to resolve blind directions in a fit by adding observables that probe complementary directions in the SMEFT parameter space. The resolution can be improved by adding observables like the energy asymmetry in top-antitop-jet production~\cite{Basan:2020btr} or observables of the top polarization like spin correlations, which are sensitive to other combinations of the two couplings.

When combining different observables in global fits, it is crucial to include correlations in both predictions and measurements. We illustrate the impact of correlations by performing two fits of the following set of observables
\beq
\sigma_{t\bar t}\,,\ \frac{1}{\sigma}\frac{d\sigma}{d m_{t\bar t}},\ \frac{1}{\sigma}\frac{d\sigma}{d\Delta y_{t\bar t}},\ A_C\,.
\eeqn
In the first fit, we assume that the systematic uncertainties in bins of differential distributions are fully correlated; in the second fit, we assume no correlations. The results in Fig.~\ref{fig:top-antitop}, right, show that the correlations relax the bounds on the Wilson coefficients, providing us with a more realistic interpretation of the data than without correlations. More details can be found in Refs.~\cite{Brivio:2019ius,Bissmann:2019qcd}.

\section{The top-bottom connection}\label{sec:top-bottom}
New physics effects in top observables at the LHC are linked to effects in flavor observables through the renormalization group. Each WET coefficient $\mathcal{C}_a$ at the bottom mass scale $m_b$ can be expressed as a linear combination of SMEFT coefficients $\{C_i\}$ at the top mass scale $m_t$,
\beq
\mathcal{C}_a(m_b) = F(C_i(m_t))\,.
\eeqn
By including flavor observables in global fits, we probe new directions in the SMEFT space and improve the sensitivity to new physics.

The relative impact of top and bottom observables in a combined analysis strongly depends on the flavor structure of the Wilson coefficients. This is due to the fact that top observables probe mostly flavor-diagonal couplings, while bottom observables are also very sensitive to flavor-changing couplings. We can exploit the top-bottom connection to probe the flavor structure of a UV theory.

\begin{figure}[t!]
\centering
\raisebox{0.3cm}{\includegraphics[width=0.43\textwidth]{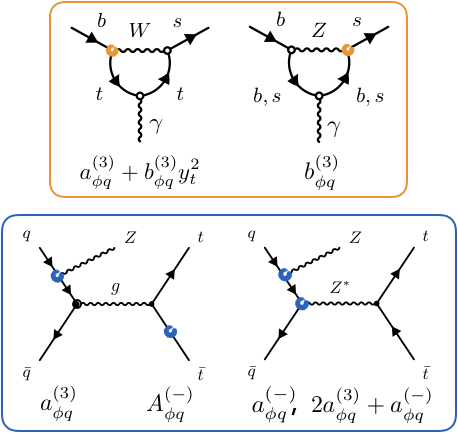}} \hspace*{0.05\textwidth}  \includegraphics[width=0.48\textwidth]{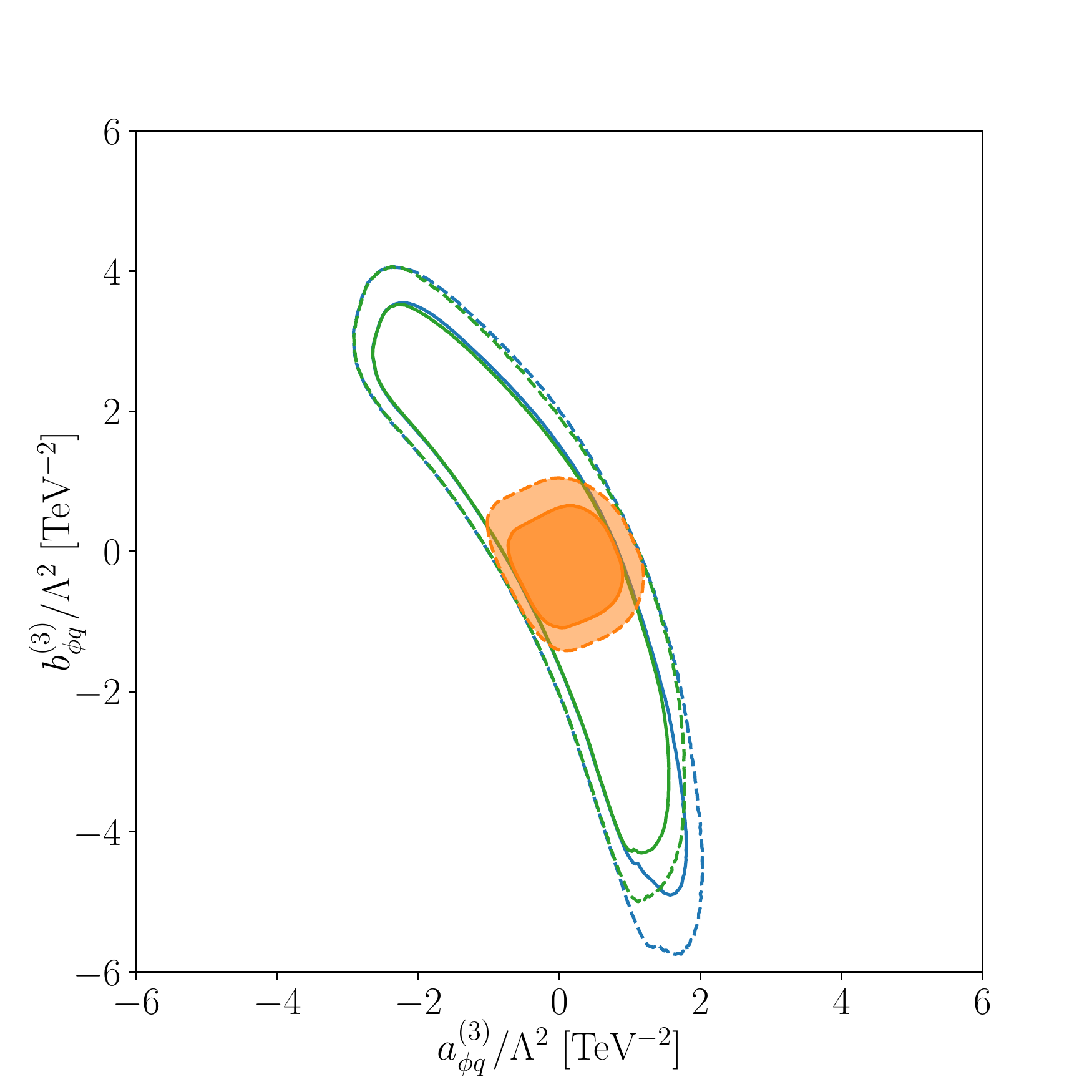}
\caption{Left: SMEFT contributions of $O_{\phi q}^{(1)}$ and $O_{\phi q}^{(3)}$ to $B\to X_s \gamma$ (top) and to $pp\to t\bar t Z$ (bottom) in minimal flavor violation. Right: bounds obtained from a combined fit of the flavor parameters $\{a_{\phi q}^{(3)},b_{\phi q}^{(3)},a_{\phi q}^{(-)},b_{\phi q}^{(-)}\}$ to top observables (blue), top \& $\mathcal{B}(B_s\to \mu^+\mu^-)$ (green) and top \& $\mathcal{B}(B_s\to \mu^+\mu^-)$ \& $\mathcal{B}(B \to X_s \gamma)$ (orange). The flavor parameters $a,b$ are defined at the top mass scale $m_t$. Details in Ref.~\cite{Bruggisser:2021duo}.}
\label{fig:top-flavor}
\end{figure}

Here we show how to resolve the flavor structure of the Wilson coefficients in the framework of minimal flavor violation. Details are given in Ref.~\cite{Bruggisser:2021duo}. We focus on two operators
\beq
O_{\phi q}^{(1)} = (H^\dagger \!\stackrel{\longleftrightarrow}{iD^\mu}\! H)(\overline{Q}^k\gamma_\mu Q^l),\qquad O_{\phi q}^{(3)} = (H^\dagger \!\stackrel{\longleftrightarrow}{iD^\mu}\! \tau^a H)(\overline{Q}^k\gamma_\mu \tau^a Q^l)\,,
\eeqn
where the two currents in $O_{\phi q}^{(1)}$ transform as singlets under weak interactions, while $O_{\phi q}^{(3)}$ has a triplet structure encoded in the $SU(2)$ generators $\tau^a$. Furthermore, $Q^{k}$ and $Q^{l}$ stand for the left-handed quark doublets of generation $k,l = \{1,2,3\}$. Assuming that the only sources of flavor symmetry breaking in the UV theory are -- as in the Standard Model -- the Yukawa couplings, we can expand the Wilson coefficients as
\beq
(C)_{kl} = \left(a\, {\bf 1} + b\, Y_U Y_U^\dagger + c\, Y_D Y_D^\dagger + \dots \right)_{kl} = a\,\delta_{kl} + b y_t^2\, \delta_{k3}\delta_{l3} \,+\, \mathcal{O}(y_b^2)\,.
\eeqn
The parameter $a$ corresponds to flavor-universal couplings, and $b$ induces flavor-breaking couplings. In Fig.~\ref{fig:top-flavor}, left, we illustrate how the flavor parameters contribute to the rare meson decay $B\to X_s\gamma$ (orange) and to hadronic $t\bar t Z$ production (blue).\footnote{The combinations $a^{(-)} = a^{(1)} - a^{(3)}$ etc. parametrize the coupling to up-type quarks, and $A = a + b\,y_t^2$ is the relevant combination for third-generation quarks.} By combining top and bottom observables, we can probe the flavor structure of UV interactions by distinguishing between flavor-universal and flavor-breaking couplings.

In Fig.~\ref{fig:top-flavor}, right, we show the bounds resulting from a combined fit of top observables, as well as the meson decays $B\to X_s \gamma$ and $B_s\to \mu^+\mu^-$. While the top fit alone (blue) leaves space for flavor breaking along $b$, the combined fit (orange) resolves this direction and suggests that flavor-breaking contributions, if present, should be small. Interestingly, this insight relies on one-loop effects of $O_{\phi q}^{(3)}$ to $B\to X_s \gamma$, which are very sensitive to flavor breaking. Notice also that electroweak contributions to $t\bar t Z$ production with a virtual $Z$ boson are important to correctly describe this process in SMEFT.

\section{Conclusions}
These two examples illustrate how to resolve the structure of Wilson coefficients in the SMEFT framework by combining top and bottom observables. The fit results are presented at the top mass scale, but can be translated to any scale $\Lambda > m_t$ by evolving the Wilson coefficients via the renormalization group, where they can be matched with any concrete UV theory that does not involve new light particles.

Much more can be learned about UV physics by strategically building on this proof of principle. Good places to look for indirect signs of new physics are observables that are precise and/or suppressed in the Standard Model. At the LHC, associated $tZ$, $tW$, $th$ production are very sensitive processes~\cite{Degrande:2018fog}, as well as tails of kinematic distributions that probe new physics at the highest available energies~\cite{Englert:2016aei,Maltoni:2019aot}. Combined fits to LHC data in the top, Higgs, and electroweak sectors~\cite{Ellis:2020unq}, as well as combinations of Higgs and electroweak observables with flavor observables~\cite{Aoude:2020dwv} are promising steps towards a truly global SMEFT analysis that will allow us to resolve more and more of what might be hiding in the UV.

\Acknowledgements
I thank the organizers of Top 2020 for inviting me and for bringing the lastest news from the top to our living rooms. This work has been supported by the DFG (German Research Foundation) under grant no. 396021762--TRR 257.


\end{document}